\documentclass[12pt,amsmath,amssymb,showpacs,showkeys]{revtex4}
\usepackage{amsmath,amsfonts,amsthm,amscd,amssymb,latexsym}
\usepackage{bm}
\usepackage{dcolumn}
\usepackage{graphicx}
\usepackage{epstopdf}
\usepackage{color}
\usepackage{epsf}
\usepackage{epsfig}
\usepackage{graphicx, epic, eepic, color}

\newcommand{\beq}{\begin{equation}}
\newcommand{\eeq}{\end{equation}}

\begin{document}

\title{General Relativistic Gravity Gradiometry}

\author{Bahram \surname{Mashhoon}}
\email{MashhoonB@missouri.edu}
\affiliation{Department of Physics and Astronomy, University of Missouri, Columbia, Missouri 65211, USA\\
and\\
School of Astronomy, Institute for Research in Fundamental
Sciences (IPM), P. O. Box 19395-5531, Tehran, Iran}

\date{\today}
\begin{abstract}
Gravity gradiometry within the framework of the general theory of relativity involves the
measurement of the elements of the relativistic tidal matrix, which is theoretically obtained via the projection of the spacetime curvature tensor upon the nonrotating orthonormal tetrad frame of a geodesic observer. The behavior of the measured components of the curvature tensor under Lorentz boosts is briefly described in connection with the existence of certain special tidal directions.  Relativistic gravity gradiometry in the exterior gravitational field of a rotating mass is discussed and a gravitomagnetic beat effect along an inclined spherical geodesic orbit is elucidated. 
\end{abstract}

\pacs{04.20.Cv, 04.25.Nx, 04.80.-y}
\keywords{relativistic gravity gradiometry, post-Schwarzschild approximation, Kerr spacetime}

\maketitle

\section{Newtonian Gravity Gradiometry}

Consider a distribution of matter of density $\rho(t, \mathbf{x})$ and the corresponding Newtonian gravitational potential $\Phi(t, \mathbf{x})$ in an inertial frame of reference. In a source-free region of space, we imagine two nearby test masses $m_a$ and $m_b$ that fall freely in the potential $\Phi$ along trajectories $\mathbf{x}_a(t)$ and $\mathbf{x}_b(t)$, respectively. Choosing one of these as the reference trajectory, we are interested in the \emph{relative} motion of these test particles. With $\mathbf{x}_b(t)$ as the fiducial path, let us define $\boldsymbol{\xi}(t) = \mathbf{x}_a(t) - \mathbf{x}_b(t)$. Newton's second law of motion implies that the instantaneous deviation vector $\boldsymbol{\xi}(t)$ between the neighboring paths satisfies the tidal equation
\begin{equation}\label{I1}
\frac{d^2\xi^i}{dt^2}+\kappa^i{}_j\, \xi^j + O (|\boldsymbol{\xi}|^2)=0\,,
\end{equation}
where $- \kappa^i{}_j\, \xi^j$ is the first-order tidal acceleration and
\begin{equation}\label{I2}
\kappa_{ij}(t, \mathbf{x})= \frac{\partial^2 \Phi}{\partial x^i\partial x^j}\,
\end{equation}
is the symmetric tidal matrix evaluated along the reference trajectory. In the Newtonian theory of gravitation, gravity gradiometry involves the measurement of $\kappa_{ij}(t, \mathbf{x})$, which is the gradient of the acceleration of gravity and can be determined, in principle, by means of Eq.~\eqref{I1}.

The tidal matrix in Eq.~\eqref{I2} is independent of the test masses $m_a$ and $m_b$  as a consequence of the principle of equivalence. The principle of equivalence of gravitational and inertial masses ensures the universality of the gravitational interaction. The modern history of the science of gravity gradiometry can be traced back to the pioneering efforts of L. E\"otv\"os, who employed a torsion-balance method to test the principle of equivalence (1889 - 1922).

In Eq.~\eqref{I2}, Poisson's equation for $\Phi$, $\nabla^2 \Phi=4 \pi G \,\rho$, reduces to Laplace's equation, $\nabla^2 \Phi=0$, in the source-free region under consideration. In this case, $\nabla^2 \kappa_{ij}=0$ and hence each element of the Newtonian tidal matrix is a \emph{harmonic} function. Moreover,  tidal matrix~\eqref{I2} is \emph{traceless}; therefore,  the shape of a tidally deformed 
test body would generally tend to either a cigar-like or a pancake-like configuration when tides are dominant, since the symmetric and traceless tidal matrix can in general have either two positive and one negative or  one positive and two negative    eigenvalues, respectively. 

In recent years, gravity gradiometers  of high sensitivity have been developed; indeed, the Paik  gravity gradiometer employs superconducting quantum interference devices~\cite{HJP1,HJP2,HJP3}.  Furthermore, gravity gradients can now be measured via atom interferometry as well~\cite{AT1, AT2}. Gravity gradiometry has many important practical applications.The magnitude of a gravity gradient is usually expressed in units of  E\"otv\"os, 1 E = $10^{-9}$ s$^{-2}$.

To extend the treatment of gravity gradiometry to the relativistic domain, it is necessary to introduce the quasi-inertial Fermi normal coordinate system that can provide a physically meaningful interpretation of the measurement of relative motion within the framework of general relativity (GR). In GR, masses $m_a$ and $m_b$ follow geodesics and a hypothetical observer comoving with the fiducial test mass $m_b$ would set up in the neighborhood of the reference trajectory a laboratory where the motion of $m_a$ could be monitored. Such a quasi-inertial frame is represented by the Fermi normal coordinate system~\cite{LC, SY, SY1}. In our treatment of Fermi coordinates in the next section, we employ an extended framework~\cite{mas77, CM, CM1, CM2}, since in practice nongravitational accelerations and rotations may be present.

\section{Fermi Coordinates}

To develop the relativistic analogs of Eqs.~\eqref{I1} and~\eqref{I2}, we consider a congruence of future-directed timelike paths representing the world lines of test masses in a gravitational field.  Next, we choose a reference path in the congruence and establish a local quasi-inertial Fermi system of geodesic coordinates in its neighborhood. This is necessary in order to provide a physically meaningful interpretation of the measurement of relative motion from the standpoint of the observer comoving with the reference test mass along the fiducial world line $\bar{x}^\mu(\tau)$.  The observer has proper time $\tau$ and carries an orthonormal  tetrad frame $\lambda^{\mu}{}_{\hat \alpha} (\tau)$ along $\bar{x}^\mu$; that is, $g_{\mu \nu} \,\lambda^\mu{}_{\hat \alpha}\,\lambda^\nu{}_{\hat \beta}= \eta_{\hat \alpha \hat \beta}$, where $g_{\mu \nu}$ is the spacetime metric and $\eta_{\hat \alpha \hat \beta}$ is the Minkowski metric given by diag$(-1, 1, 1, 1)$ in our convention.  Here, $\lambda^{\mu}{}_{\hat 0} (\tau) = d\bar{x}^\mu/d\tau$ is the observer's temporal axis and its local frame is carried along its path according to
\begin{equation}\label{II1}
\frac{D\lambda^{\mu}{}_{\hat \alpha}}{d\tau} = \phi_{\hat \alpha}{}^{\hat \beta} \,\lambda^{\mu}{}_{\hat \beta}\,,
\end{equation}
where $\phi_{\hat \alpha \hat \beta}$ is the observer's antisymmetric acceleration tensor. Greek indices run from $0$ to $3$, while Latin indices run from $1$ to $3$. The signature of the spacetime metric is $+2$ and units are chosen such that $c = G = 1$, unless specified otherwise.

In close analogy with the electromagnetic field tensor, we can decompose the acceleration tensor into its ``electric" and ``magnetic" components, namely, $\phi_{\hat \alpha \hat \beta} \to (-\mathbf{A}, \boldsymbol{\Omega})$, where $\mathbf{A}(\tau)$ is a spacetime scalar that represents the translational acceleration of the fiducial observer and $\boldsymbol{\Omega}(\tau)$ is a spacetime scalar that represents its rotational acceleration. More precisely, the reference observer in general follows an accelerated world line with 
\begin{equation}\label{II2}
\frac{d^2\bar{x}^{\mu}}{d\tau^2}+\Gamma^\mu_{\nu \sigma}\,\frac{d\bar{x}^\nu}{d\tau}\,\frac{d\bar{x}^\sigma}{d\tau} = \mathcal{A}^\mu\,,
\end{equation}
where
\begin{equation}\label{II3}
\mathcal{A}^\mu = A^{\hat i}\,\lambda^{\mu}{}_{\hat i}\,
\end{equation}
and $\boldsymbol{\Omega}$ is the angular velocity of the rotation of the observer's spatial frame with respect to a locally nonrotating (i.e. Fermi-Walker transported) frame. 

At each event $\bar{x}^\mu(\tau)$ along the reference world line, we imagine all spacelike geodesic curves that start out from this event and are normal to the reference world line. These generate a local hypersurface. Let $x^\mu$ be an event on this hypersurface sufficiently close to the reference world line such that there is a \emph{unique} spacelike geodesic of proper length $\sigma$ that connects $\bar{x}^\mu(\tau)$ to $x^\mu$.  We define $\xi^\mu$ to be a unit spacelike vector that is tangent to the unique spacelike geodesic
 at $\bar{x}^\mu(\tau)$, so that  $\xi_\mu(\tau)\,\lambda^{\mu}{}_{\hat 0}(\tau) = 0$. Then, to event $x^\mu$ one assigns Fermi coordinates $X^{\hat \mu}$, where
\begin{equation}\label{II4}
X^{\hat 0} := \tau\,, \qquad X^{\hat i} := \sigma\, \xi^\mu(\tau)\, \lambda_{\mu}{}^{\hat i}(\tau)\,.
\end{equation}
The reference observer has Fermi coordinates $X^{\hat \mu} = (\tau, 0, 0, 0)$ and  is thus permanently fixed at the spatial origin of the Fermi coordinate system. Henceforth, we find it convenient to express Fermi coordinates as
$X^{\hat \mu} = (T, \mathbf{X})$, where $|\mathbf{X}|= \sigma$. When $\sigma \ne 0$,  $X^{\hat i}/\sigma = \xi^\mu(\tau)\, \lambda_{\mu}{}^{\hat i}(\tau)$, for $i = 1, 2, 3$, are the corresponding direction cosines at proper time $\tau$ along $\bar{x}^\mu$. 

The Fermi coordinate system is admissible in a cylindrical domain along $\bar{x}^\mu$ of radius $|\mathbf{X}| \sim \mathfrak{R}$, where $\mathfrak{R}$ is a certain minimal radius of curvature of spacetime along the reference world line. 

The spacetime metric in Fermi coordinates is given by
\beq \label{II4a}
 ds^2= g_{\hat \mu \hat \nu}(T, \mathbf{X})\,dX^{\hat \mu} \,dX^{\hat \nu}\,, 
\eeq
 where
\begin{eqnarray}\label{II5}
g_{\hat 0 \hat 0} &=& -P^2 + Q^2  - R_{\hat 0 \hat i \hat 0 \hat j}\,X^{\hat i}\,X^{\hat j} + O(|\mathbf{X}|^3)\,,\\
g_{\hat 0 \hat i} &=& Q_{\hat i} -\frac{2}{3} \,R_{\hat 0 \hat j \hat i \hat k}\,X^{\hat j}\,X^{\hat k} + O(|\mathbf{X}|^3)\,,\label{II6}\\
g_{\hat i \hat j} &=& \delta_{\hat i \hat j} -\frac{1}{3} \,R_{\hat i \hat k \hat j \hat l}\,X^{\hat k}\,X^{\hat l} + O(|\mathbf{X}|^3)\,.\label{II7}
\end{eqnarray}
Here,  $P$ and $\mathbf{Q}$,
\begin{equation}\label{II8}
P := 1 + \mathbf{A}(T) \cdot \mathbf{X}\,, \qquad \mathbf{Q} := \boldsymbol{\Omega}(T) \times \mathbf{X}\,,
\end{equation}
are related to the local translational and rotational accelerations of the reference observer, respectively,  and
\begin{equation}\label{II9}
R_{\hat \alpha \hat \beta \hat \gamma \hat \delta}(T) := R_{\mu \nu \rho \sigma}\,\lambda^{\mu}{}_{\hat \alpha}\,
\lambda^{\nu}{}_{\hat \beta}\,\lambda^{\rho}{}_{\hat \gamma}\,\lambda^{\sigma}{}_{\hat \delta}
\end{equation}
is the projection of the Riemann curvature tensor along $\bar{x}^\mu$ on the  tetrad frame of the reference observer. 

Fermi coordinates are invariantly defined and can have advantages over other physically motivated coordinate systems such as radar coordinates~\cite{Bini:2004qj}; therefore, they have been applied in many different contexts. For instance, Fermi coordinates have been employed to elucidate dynamics of astrophysical jets~\cite{MaMc, CM3, CM4, CM5, CM6}.

\section{Relativistic Gravity Gradiometry}

In Einstein's GR, gravity gradiometry involves the measurement of the gravitational field, which is represented by the Riemannian curvature of spacetime. When an observer measures a  gravitational field, the curvature tensor must be projected onto the tetrad frame of the observer. 

It is now straightforward to express the equation of motion of any other test mass in the Fermi coordinate system and study the motion of the test mass relative to the fiducial test mass that follows  world line $\bar{x}^\mu$. This general framework is necessary in practice, since the motion of the reference test mass may involve translational and rotational accelerations of nongravitational origin. These are absent, however, in the ideal case of purely tidal relative motion. To illustrate this ideal situation, let us assume that $\phi_{\hat \alpha \hat \beta}=0$, so that the reference path $\bar{x}^\mu$ is a timelike geodesic and the orthonormal tetrad frame is parallel transported along the fiducial geodesic world line, i.e. $D\lambda^\mu{}_{\hat \alpha}/d\tau=0$.  The geodesic equation of  motion of a free test particle in the corresponding Fermi coordinates relative to the reference test mass that is fixed at the spatial origin 
of Fermi coordinates can be expressed in terms of relative separation $\mathbf{X}$ as
\begin{eqnarray}\label{II10} 
 \frac{d^2X^{\hat i}}{dT^2}&+&R_{\hat 0\hat i\hat 0\hat j}X^{\hat j}+2\,R_{\hat i\hat k\hat j\hat 0}V^{\hat k}X^{\hat j}\nonumber \\
 &+&\frac{2}{3}\,\left(3R_{\hat 0\hat k\hat j\hat 0}V^{\hat i}V^{\hat k}
+R_{\hat i\hat k\hat j\hat l}V^{\hat k}V^{\hat l}+ R_{\hat 0\hat k\hat j\hat l}V^{\hat i}V^{\hat k}V^{\hat l}\right) X^{\hat j} +~ O(|\mathbf{X}|^2) = 0\,.
\end{eqnarray}
This geodesic deviation equation is a generalized Jacobi equation
\cite{CM} in which the rate of geodesic separation (i.e.  the relative 
velocity of the test particle) ${\mathbf V}=d{\mathbf X}/dT$ is
in general arbitrary; however, $|{\mathbf V}|<1$ at $\mathbf{X} = 0$.  It is clear from Eq.~\eqref{II10} that \emph{all} of the curvature components in Eq.~\eqref{II9} can be measured from a careful study of the motion of the test masses in the congruence relative to the fiducial observer. Neglecting terms  in the  relative velocity $\mathbf V$, Eq.~\eqref{II10}  reduces to the Jacobi equation, 
\begin{equation}\label{II11} 
\frac{d^2{X^{\hat i}}}{dT^2} +\mathcal{K}^{\hat i}{}_{\hat j} X^{\hat j} + O(|\mathbf{X}|^2) = 0\,,
\end{equation}
which is the relativistic analog of the Newtonian tidal equation given by Eq.~\eqref{I1}, and
\begin{equation}\label{II12} 
\mathcal{K}_{\hat i \hat j}=R_{\hat 0 \hat i \hat 0 \hat j}\,.
\end{equation}
This symmetric relativistic tidal matrix is traceless in Ricci-flat regions of spacetime and reduces in the nonrelativistic limit to the Newtonian tidal matrix~\eqref{I2}. 

The relativistic tidal matrix is thus determined by the projection of the Riemann curvature tensor upon the parallel-transported tetrad frame of the fiducial geodesic observer. The local spatial frame of the fiducial observer is unique up to a \emph{constant} spatial rotation corresponding to the choice of the initial orthonormal triad in Eq.~\eqref{II1}. The freedom in the choice of the initial local triad implies that the form of the tidal matrix is unique up to a constant spatial rotation.  

The Jacobi equation can be used to study the influence of a gravitational field on the relative motion of nearby test masses in general relativity~\cite{Pir, PZ, Podolsky:2012he, CH}. Einstein's field equations locally relate the energy-momentum tensor of matter to the Ricci tensor.  At any event in spacetime, the Riemann curvature tensor can be decomposed into a matter part and a part that is independent of matter; that is, 
\begin{equation}\label{II13}
R_{\mu \nu \rho \sigma} = C_{\mu \nu \rho \sigma} +g_{\mu[\rho}\,R_{\sigma]\nu} - g_{\nu[\rho}\,R_{\sigma]\mu} - \frac{1}{6}\,(g_{\mu \rho}\,g_{\nu \sigma} - g_{\mu \sigma}\,g_{\nu \rho})\,R\,,
\end{equation}
where $C_{\mu \nu \rho \sigma}$ is the  traceless Weyl curvature tensor that represents the ``free" gravitational field. At any point on the manifold, the Riemann tensor has in general 20 independent components, whereas the Ricci tensor has 10 independent components. Beyond any point on the spacetime manifold, the two parts of the curvature tensor are connected to each other via the Bianchi identity $R_{\mu \nu[\rho \sigma; \delta]} = 0$. Introducing decomposition~\eqref{II13} into the Jacobi equation and employing a canonical null tetrad frame, Szekeres has shown via the Petrov classification that the behavior of the free part of the gravitational field can be described in terms of the superposition of a transverse wave component, a longitudinal component and a Coulomb component~\cite{PZ}. The matter part has been treated in~\cite{CH}. Some of the basic astrophysical applications of Eq.~\eqref{II11} have been studied in~\cite{mas77, mas73, mas75}.

The Gravity Probe B (``GP-B") experiment has recently measured the exterior gravitomagnetic field of the Earth~\cite{Ever}. The gravitomagnetic field of a rotating mass contributes to the spacetime curvature and can thus influence the relative tidal motion of nearby test masses. In 1980, Braginsky and Polnarev~\cite{BaPo} proposed an experiment to measure such an effect in a space platform in orbit around the Earth, since they claimed that such an approach could  
circumvent many of the difficulties associated with the GP-B experiment. However, in 1982, Mashhoon and Theiss~\cite{MaTh} showed that to measure the relativistic rotation-dependent tidal acceleration in a space platform, the local gyroscopes that would fix the local spatial frame carried by the space platform must satisfy the same performance criteria as in the GP-B experiment. 

The achievements of the GP-B  could possibly be integrated with Paik's superconducting gravity gradiometer~\cite{Paik} in future space experiments in order to measure the tidal influence of the gravitomagnetic field using an orbiting platform~\cite{PMW, MPW}.  We will consider the prediction of GR for the nature of the tidal matrix in such experiments in Section V.

\section{Special Tidal Directions} 

Let us return to the main focus of relativistic gravity gradiometry, namely, the determination of the Riemann curvature tensor projected on the tetrad frame of the fiducial observer as in Eq.~\eqref{II9}. Taking advantage of the symmetries of the Riemann tensor, this quantity can be represented by a $6\times6$ matrix $\mathcal{R} = (\mathcal{R}_{IJ})$, where the indices $I$ and $J$ range over the set $(01,02,03,23, 31,12)$. Thus we can write
\begin{equation}
\label{T1}
\mathcal{R}=\left[
\begin{array}{cc}
\mathcal{E} & \mathcal{B}\cr
\mathcal{B^{\dagger}} & \mathcal{S}\cr
\end{array}
\right]\,,
\end{equation}
where $\mathcal{E}$  and $\mathcal{S}$ are symmetric $3\times3$ matrices and  $\mathcal{B}$ is traceless. The tidal matrix $\mathcal{E}$ represents the ``electric" components of the curvature tensor as measured by the fiducial observer, whereas $\mathcal{B}$ and $\mathcal{S}$ represent its ``magnetic" and ``spatial" components, respectively.  Imagine next an observer that is boosted with speed $\beta$ in a given direction with respect to the fiducial observer at the same event in spacetime. Let $\mathcal{R'}$ be the Riemann curvature tensor as measured by the boosted observer. It turns out that under the boost the elements of $\mathcal{E}$, $\mathcal{B}$ and $\mathcal{S}$ in the direction parallel to the direction of the boost are \emph{not} affected, whereas those perpendicular to the direction of the boost are enhanced by $\gamma^2$,  where $\gamma = (1-\beta^2)^{-1/2}$ is the Lorentz factor; moreover, the mixed elements are enhanced by a factor of $\gamma$. This circumstance is reminiscent of the behavior of the electromagnetic field under a boost: The components of the electric field ($E$) and magnetic field ($B$) parallel to the direction of the boost remain the same as before, while those perpendicular to the direction of the boost are enhanced by a factor of $\gamma$. 

In this way the strength of the gravitational field can be augmented by a factor of $\gamma^2$; alternatively, one can say that the radius of curvature of spacetime measured by the boosted observer is Lorentz contracted~\cite{BM1, BM2}. In Ricci-flat regions of spacetime, Eq.~\eqref{T1} simplifies, since   
$\mathcal{S} = -\mathcal{E}$, $\mathcal{E}$ is traceless and $\mathcal{B}$ is symmetric. Hence, the Weyl curvature tensor with 10 independent components is completely determined by its ``electric" and ``magnetic" components that are symmetric and traceless $3\times3$ matrices. 

These results imply that a gravity gradiometer would in general measure extremely strong tidal forces when it moves very fast ($\beta \to 1$). However, along certain exceptional directions in space, such as the radial direction in the exterior Schwarzschild spacetime, tidal forces remain finite as $\beta \to 1$~\cite{BM1,BM2}. Along such a \emph{special tidal direction}, the corresponding world line of the boosted observer approaches a null direction in the local null cone as $\beta \to 1$. In this way, special tidal directions are associated with certain tidally \emph{nondestructive null directions} in spacetime. The significance of these null directions can be further elucidated via the invariant Petrov classification of gravitational fields. 

The Petrov classification involves the Weyl curvature tensor and provides an invariant characterization of a gravitational field. This can be accomplished, for instance, in terms of the \emph{principal null directions} of the Weyl tensor. A vector $k$, $k_\alpha\,k^\alpha = 0$, which satisfies the condition
\begin{equation}
\label{T2}
k_{[\alpha}\,C_{\mu]\nu\rho[\sigma}\,k_{\beta]}\,k^\nu\,k^\rho = 0\,
\end{equation}
is a principal null direction of the Weyl tensor. In a gravitational field, at least one and at most four such null vectors exist at each event in spacetime~\cite{R1, Griffiths:2009dfa}.

The basic mathematical connection between the special tidal directions and the principal null directions of the Weyl tensor has been established by Beem and Parker~\cite{BePa} and Hall and Hossack~\cite{HaHo}. It turns out that in general a nondestructive null direction at a point $p$ in spacetime is a principal null direction of the Weyl tensor at $p$; moreover, it is a \emph{repeated} principal null direction of the Weyl tensor at $p$ if and only if it is a Ricci eigendirection at $p$. A vector $N^\mu$ is a Ricci eigendirection at $p$ if
\begin{equation}
\label{T3}
R_{\mu\nu}\,N^\nu = \sigma \,N_\mu\,
\end{equation}
for a real number $\sigma$ at $p$. This means that in a Ricci-flat spacetime, or more generally when 
\begin{equation}
\label{T4}
R_{\mu\nu} = \Lambda\,g_{\mu \nu}\,
\end{equation}
for a real number $\Lambda$, a special tidal direction at $p$ corresponds to a \emph{repeated principal null direction of the Weyl tensor} at $p$. A vector $N^\mu$ is a repeated principal null vector of the Weyl tensor at $p$ if $N_\alpha\,N^\alpha = 0$ and 
\begin{equation}
\label{T4}
C_{\mu\nu\rho\sigma} \,N^\nu\,N^\sigma = \lambda\,N_{\mu}\,N_{\rho}\,
\end{equation}
at $p$ for a real number $\lambda$. There are at least zero and at most two such directions at each event in Ricci-flat spacetimes. 

Let us assume that the Weyl tensor vanishes at $p$, then a nondestructive null direction at $p$ exists if and only if it is a Ricci eigendirection at $p$. In this case, one can have $0, 1, 2$ or $\infty$ nondestructive null directions at $p$~\cite{HaHo}. For instance, there are no special tidal directions in any of the standard Friedmann-Lema\^{\i}tre-Robertson-Walker cosmological models. However, every direction is a special tidal direction in a spacetime of constant nonzero curvature, namely, de Sitter (or anti-de Sitter) universe. 

The behavior of the measured components of the Riemann curvature tensor under boosts along special tidal directions can be determined based on the results given in Ref.~\cite{BM2}. Let us consider, in particular, the Kerr gravitational field, which is of type \emph{D} in the Petrov classification. The Weyl tensor at each point in this spacetime has two repeated principal null directions; therefore, there are two special tidal directions at each event. For example, along the axis of rotation, the outgoing and ingoing radial directions are the special tidal directions, see Ref.~\cite{MaMc} for an extended treatment. In general, along the special tidal directions in Kerr spacetime, the curvature remains invariant under boosts ($\mathcal{R'} = \mathcal{R}$); in fact, the ``electric" and ``magnetic" components of the curvature can be made ``parallel" such that the super-Poynting vector
\begin{equation}
\label{T5}
\mathcal{P}_{\hat i}  = -\epsilon_{\hat i \hat j \hat k}\,(\mathcal{E}\mathcal{B})_{\hat j \hat k}\,
\end{equation} 
vanishes. An analogous situation is encountered in the case of the electromagnetic field in an inertial frame in Minkowski spacetime. If the electromagnetic field is not \emph{null}, so that the invariants $E^2-B^2$ and $\mathbf{E}\cdot \mathbf{B}$ do not simultaneously vanish, then a boost with velocity $v$ along the Poynting vector, i.e.
\begin{equation}
\label{T6}
\frac{\mathbf{v}}{1+v^2} = \frac{\mathbf{E}\times \mathbf{B}}{E^2+B^2}\,,
\end{equation} 
renders the electric and magnetic fields parallel in the boosted frame. In the new inertial frame, the Poynting vector vanishes and any boost along the common direction of the fields leaves them invariant. The analogy between the electromagnetic field and algebraically special gravitational fields of types \emph{D} and \emph{N} has been treated in Ref.~\cite{BM2}.

\section{Tidal Matrix Around A Rotating Mass}

To get some idea regarding the form of the relativistic tidal matrix, it is instructive to consider first the tidal field along stable circular orbits in the equatorial plane of the Kerr spacetime. 
The exterior Kerr metric can be expressed as~\cite{Chandra}
\begin{equation}\label{K1}
ds^2=-dt^2+\frac{\Sigma}{\Delta}dr^2+\Sigma\, d\theta^2 +(r^2+a^2)\sin^2\theta\, d\varphi^2+\frac{2Mr}{\Sigma}(dt-a\sin^2\theta\, d\varphi)^2\,,
\end{equation}
where $M$ is the mass of the gravitational source,  $a=J/M$ is the specific angular momentum of the source, $(t,r,\theta,\varphi)$ are the standard Boyer-Lindquist coordinates and
\beq\label{K2}
\Sigma=r^2+a^2\cos^2\theta\,,\qquad \Delta=r^2-2Mr+a^2\,.
\eeq
The Kerr metric contains  dimensionless gravitoelectric and gravitomagnetic potentials $\mathcal{U} = GM/(c^2r)$ and $\mathcal{V}= GJ/(c^3\,r^2)$, which correspond to the mass and angular momentum of the source, respectively. For instance, in the case of the Earth, we have $\mathcal{U}_{\oplus} \approx 6\times 10^{-10}$ and $\mathcal{V}_{\oplus} \approx 4\times 10^{-16}$. 

We are interested in the tidal matrix along the circular equatorial trajectory of a fiducial test mass that follows  a future-directed timelike geodesic world line about the Kerr source. The circular orbit has a fixed radial coordinate $r_0$ and orbital frequency~\cite{Chandra}
\beq\label{K2a}
\frac{d\varphi}{d\tau} =  \frac{\omega_0}{(1-3\frac{M}{r_0} + 2 a \omega_0)^{1/2}}\,,
\eeq
where the Keplerian frequency $\omega_0$ is given by
\beq\label{K2b}
\omega_0^2 = \frac{M}{r_0^3}\,.
\eeq
The circular geodesic orbit is such that at proper time $\tau =0$, the azimuthal coordinate vanishes (i.e. $\varphi =0$). Moreover, at this event, the initial directions of the orthonormal triad $\lambda^\mu{}_{\hat i}$, $i = 1, 2, 3$,  point along the spherical polar coordinate directions. The spatial triad then undergoes parallel propagation along the circular orbit. The resulting  radial and tangential components of the spatial frame, namely, $\lambda^\mu{}_{\hat 1}$ and  $\lambda^\mu{}_{\hat 3}$, respectively, turn out to be periodic in $\tau$ with period $2\pi/\omega_0$. The difference between the orbital frequency~\eqref{K2a} and the Keplerian frequency $\omega_0$ leads to a combination of prograde geodetic and retrograde gravitomagnetic precessions of these frame components with respect to static inertial observers at spatial infinity in the asymptotically flat Kerr spacetime~\cite{Bini:2016xqg}. 

The tidal matrix is obtained as a certain symmetric and traceless projection of the Riemann curvature tensor evaluated along the orbit. The nonzero components of the tidal matrix consist of constant terms proportional to $\omega_0^2$ plus terms that are periodic in $\tau$ with frequency $2\,\omega_0$ and can be expressed as~\cite{Bini:2016xqg}
\begin{eqnarray}\label{K3}
\mathcal{K}_{\hat 1\hat 1} &=& \omega_0^2 \,[1-3\gamma_0^2\, \cos^2(\omega_0 \tau)]\,,\\
\label{K4}
\mathcal{K}_{\hat 1\hat 3} &=& \mathcal{K}_{\hat 3\hat 1} = - \frac{3}{2}\, \omega_0^2 \gamma_0^2 \, \sin(2\,\omega_0 \tau)\,,\\
\label{K5}
\mathcal{K}_{\hat 2\hat 2} &=& \omega_0^2\, (3\gamma_0^2 - 2)\,,\\
\label{K6}
\mathcal{K}_{\hat 3\hat 3} &=& \omega_0^2\, [1-3\gamma_0^2\, \sin^2(\omega_0 \tau)]\,,
\end{eqnarray}
where  $\gamma_0$ is given by
\begin{equation}\label{K7}
\gamma_0 = \left(\frac{r_0^2-2Mr_0 + a^2}{r_0^2- 3Mr_0 + 2r_0^2a\omega_0}\right)^{1/2}\,.
\end{equation}
More generally, the tidal matrix for arbitrary timelike geodesics of Kerr spacetime has been calculated by Marck~\cite{Marck}. 

Let us next consider the tidal field along a tilted spherical orbit of fixed radial coordinate $r_0$ about a \emph{slowly} rotating spherical mass. The exterior gravitational field is represented by the Kerr metric linearized in the angular momentum parameter $a$ or, equivalently, the Schwarzschild metric plus the Thirring-Lense term. The symmetric and traceless tidal matrix can be obtained from~\cite{Bini:2016xqg}
\begin{eqnarray}\label{K8}
\mathcal{K}_{\hat 1\hat 1}&=& \omega_0^2\,[1-3\,\Gamma^2\, \cos^2 (\omega_0\tau )]\,,\nonumber\\
\mathcal{K}_{\hat 1 \hat 2}&=&\mathcal{K}_{\hat 2 \hat 1} = \omega_0^2\,\Xi\, \cos (\omega_0 \tau)\,,  \nonumber\\
\mathcal{K}_{\hat 1 \hat 3}&=&\mathcal{K}_{\hat 3 \hat 1} =  - \frac{3}{2}\, \omega_0^2\,\Gamma^2\,\sin (2\,\omega_0 \tau)\,,  \nonumber\\
\mathcal{K}_{\hat 2 \hat 2}&=& \omega_0^2\,(3\,\Gamma^2-2)\,, \nonumber\\
\mathcal{K}_{\hat 2 \hat 3}&=&\mathcal{K}_{\hat 3 \hat 2} =  \omega_0^2\,\Xi\,\sin (\omega_0 \tau)\,,  \nonumber\\
\mathcal{K}_{\hat 3 \hat 3}&=&  \omega_0^2\,[1-3\,\Gamma^2\, \sin^2 (\omega_0\tau )]\,,
\end{eqnarray}
where $\Gamma$ and $\Xi$ are given by
\beq\label{K9}
\Gamma := \left(\frac{1-2\frac{M}{r_0}}{1-3\frac{M}{r_0}}\right)^{1/2}\,\left(1-\frac{a\,\omega_0\,\cos \alpha}{1-3\frac{M}{r_0}}\right)\,
\eeq
and 
\beq\label{K10}
\Xi := -3\,\frac{J}{M\,r_0^2\,\omega_0} \,\frac{\left(1-2\frac{M}{r_0}\right)^{1/2} (1+2\frac{M}{r_0})}{1-3\frac{M}{r_0}}\,\sin \alpha \,\sin \eta\,.
\eeq
Here, the angle $\alpha$ denotes the inclination of the orbit with respect to the equatorial plane and  $\eta$,
\begin{equation}\label{K11}
\eta:= \omega\tau +\eta_0\,, \qquad \omega =\frac{\omega_0}{(1-3\frac{M}{r_0})^{1/2}}\,,
\end{equation}
is the angular position of the reference test mass in the orbital plane measured from the line of the ascending node and $\eta_0$ is a constant angle.   For  $\alpha=0$, the spherical orbit under consideration turns into the circular equatorial orbit, $\Xi=0$, $\Gamma$ reduces at the linear order in $a$ to $\gamma_0$ and the tidal matrix agrees  to first order in $a$ with our previous results for the equatorial circular orbit in Kerr spacetime.

\section{Beat Effect}

The off-diagonal terms $\mathcal{K}_{\hat 1 \hat 2}=\mathcal{K}_{\hat 2 \hat 1}$ and $\mathcal{K}_{\hat 2 \hat 3}=\mathcal{K}_{\hat 3 \hat 2}$ in the tidal matrix~\eqref{K8}  represent the \emph{beat phenomenon} first pointed out in Ref.~\cite{MaTh}. The beat effect involves frequencies $\omega$ and $\omega_0$ with a beat frequency $\omega_F := \omega -\omega_0$. This is the frequency of the gravitoelectric (geodetic) Fokker precession of an ideal test gyro following a circular orbit about a spherical mass $M$. The tidal terms under consideration here that involve $\Xi$ have dominant amplitudes that are proportional to the angular momentum $J$ and are independent of the speed of light $c$.

In the work of Mashhoon and Theiss~\cite{MaTh, Mash, Theiss, T1, MA}, the resonance effect
 involving  $\omega$ and $\omega_0$ appeared in the calculation of the parallel-transported frame along the tilted spherical orbit about a rotating mass.  It resulted in a small divisor phenomenon involving $\omega_F$. For a near-Earth orbit, the Fokker period $2\pi/\omega_F$ is about $10^5$ years; therefore, in practice the Mashhoon-Theiss effect shows up as a secular term in the corresponding off-diagonal elements of the tidal matrix with amplitude 
\begin{equation}\label{M1}
\frac{9}{2}\, \frac{GJ}{c^2\,r_0^3}\,\omega_0^2\,  \tau\,\sin\alpha\,,
\end{equation}
which is consistent with the first post-Newtonian gravitomagnetic precession of the spatial frame~\cite{MA, BS}. The possibility of measuring the Mashhoon-Theiss effect via neutron interferometry~\cite{RaWe} has been discussed by Anandan~\cite{Anan}.

In the first post-Newtonian approximation, the motion of an ideal test gyro of spin $\mathbf{S}$ in orbit about a central rotating mass can be written as~\cite{Ever}
\beq\label{M2}
\frac{d\mathbf{S}}{d\tau} = (\boldsymbol{\omega}_{ge} + \boldsymbol{\omega}_{gm})\times \mathbf{S}\,,
\eeq
where 
\beq\label{M3}
\boldsymbol{\omega}_{ge} = \frac{3}{2}\,\frac{GM}{c^2\,r^3}\, \boldsymbol{\ell}\,, \qquad  \boldsymbol{\omega}_{gm} = \frac{G}{c^2\,r^5}[3\,(\mathbf{J} \cdot \mathbf{x})\,\mathbf{x}-\mathbf{J}\,r^2]\,,
\eeq
 $|\mathbf{x}|=r$ and $\boldsymbol{\ell} = \mathbf{x}\times \mathbf{v}$ is the specific angular momentum of the gyro orbit. Here, $\boldsymbol{\omega}_{ge}$ is the (gravitoelectric) \emph{geodetic} precession frequency of the gyroscope, while $\boldsymbol{\omega}_{gm}$ is its \emph{gravitomagnetic} precession frequency.  For the Earth, these precession frequencies have  been directly measured via GP-B~\cite{Ever}, which involved four superconducting gyroscopes and a telescope that were launched on 20 April 2004 into a polar Earth orbit of radius 642 km aboard a drag-free satellite.

During a satellite gradiometry experiment over a period of time $\tau$, we expect that the spatial frame of the gradiometer would accumulate geodetic and gravitomagnetic precession angles of order 
$GM\, \omega_0 \tau/(c^2\,r_0)$ and  $GJ\, \tau/(c^2\,r_0^3)$, respectively. From the comparison of these angles with Eq.~\eqref{M1}, it is clear that only the post-Newtonian gravitomagnetic secular term survives in the calculation of the projection of the Riemann tensor onto the tetrad frame of the gradiometer for the case of the tilted spherical orbit. A recent detailed discussion of the beat effect is contained in Ref.~\cite{Bini:2016xqg}, which should be consulted for a more complete treatment of relativistic gravity gradiometry in Kerr spacetime. 

The results presented in the last two sections may be considered surprising and contrary to expectations. That is, it may appear on the basis of Eq.~\eqref{II12} that the main post-Newtonian terms in $(\mathcal{K}_{\hat i \hat j})$ can be obtained intuitively by  combining Newtonian tides with the post-Newtonian motion of the spatial frame of the fiducial observer. However, in practice the projection of the Riemann tensor onto the frame of the fiducial observer involves detailed calculations in which the symmetries of the Riemann tensor need to be carefully taken into account.  

Finally, the results presented here can be used to find the main relativistic effects in the motion of the Moon. Consider the nearly circular orbit of the Earth-Moon system about the Sun. In the Fermi normal coordinate system established along this orbit, the solar tidal acceleration $- \mathcal{K}_{\hat i \hat j}\,X^{\hat j}$ is a small perturbation on the dynamics of the Earth-Moon system. Here  $\mathcal{K}_{\hat i \hat j}$ is essentially given by Eq.~\eqref{K8} and we recall that the ecliptic has a small inclination of $\alpha \approx 0.1$ with respect to the equatorial plane of the Sun. In this way, the main relativistic tidal effects in the motion of the Moon relative to the Earth caused by the gravitational field of the Sun have been determined~\cite{MaTh1, MaTh2, MaTh3}.

\section{Post-Schwarzschild Approximation}

The exterior vacuum field of a spherically symmetry mass can be uniquely described by the Schwarzschild spacetime. Small deviations from spherical symmetry can then be treated in  the post-Schwarzschild approximation scheme. This method was employed by Mashhoon and Theiss in their investigation of the relativistic tidal matrix  for a gradiometer in orbit about a rotating mass~\cite{MaTh, Mash, Theiss, T1, MA}. Thus in the first post-Schwarzschild approximation, the angular momentum of the central body is considered to first order while its mass is taken into account to all orders. The post-Schwarzschild method can be extended to include the quadrupole  and higher moments of the central body. Indeed, the effect of oblateness, treated as a first-order static deformation of the source, has been investigated by Dietmar Theiss for a gravity gradiometer on a circular geodesic orbit of small inclination about a central oblate body~\cite{Theiss,T2}.

\section{Discussion}

Gravity gradiometry in GR involves the measurement of a certain projection of the Riemannian curvature tensor of spacetime upon the orthonormal tetrad frame of an observer.  In a satellite gravity gradiometry experiment in Earth orbit, the mass $M_\oplus$, angular momentum $J_\oplus$, quadrupole moment $Q_\oplus$ and higher moments of the Earth will all contribute to the result of the experiment. For an inclined spherical geodesic orbit about a slowly rotating mass, Eq.~\eqref{K8} gives the relativistic tidal matrix to all orders in the mass of the source $M$ and to linear order in its angular momentum $J$. The result contains the beat phenomenon first pointed out by Mashhoon and Theiss~\cite{MaTh, Mash, Theiss, T1, MA}.  The corresponding influence of the quadrupole moment of the source has been studied by Theiss~\cite{Theiss,T2}.

\end{document}